\documentclass[showpacs,aps,graphicx,twocolumn]{revtex4}
\usepackage{amssymb}
\usepackage{graphicx}

\begin{document}
\title{Opaque Attack on Three-Party Quantum Secret Sharing Based on Entanglement}
\author{ Fu-Guo Deng,$^{1,2,3}$ Xi-Han Li,$^{1,2}$
and Hong-Yu Zhou$^{1,2,3}$}
\address{$^1$ Key Laboratory of Beam Technology and Material
Modification of Ministry of Education, Beijing Normal University,
Beijing 100875,
People's Republic of China\\
$^2$ Institute of Low Energy Nuclear Physics, and Department of
Material Science and Engineering, Beijing Normal University,
Beijing 100875, People's Republic of China\\
$^3$ Beijing Radiation Center, Beijing 100875,  People's Republic of
China}
\date{\today }

%That is, the dishonest agent can eavesdrop
%fully and freely in almost all the three-party QSS schemes based on
%entanglement if the quantum channel induces 50\% losses and a
%non-standard key sifting technique is used.

\begin{abstract}
Security of the three-party quantum secret sharing (QSS) schemes
based on entanglement and a collective eavesdropping check is
analyzed in the case of considerable quantum channel losses. An
opaque attack scheme is presented for the dishonest agent to
eavesdrop the message obtained by the other agent freely, which
reveals that these QSS schemes are insecure for transmission
efficiencies lower than 50\%, especially when they are used to share
an unknown quantum state. Finally, we present a general way to
improve the security of QSS schemes for sharing not only a private
key but also an unknown quantum state.

\end{abstract}
 \maketitle

In a secret sharing \cite{Blakley}, a sender, say Alice, has two
agents, Bob and Charlie who are at remote places. Alice hopes that
the agents can carry out her instruction (the message $M_A$), but
she suspects that one of them (not more than one) may be dishonest
and the dishonest one will do harm to her belongings if  he can deal
with it independently. Moreover, Alice does not know who the
dishonest one is. For the security of the secret message $M_A$,
Alice splits it into two pieces, $M_B$ and $M_C$, and then sends
them to Bob and Charlie, respectively. The two agents can read out
the message $M_A=M_B\oplus M_C$ if and only if they cooperate,
otherwise none can obtain a useful information. Quantum secret
sharing (QSS) is the generalization of classical secret sharing into
quantum scenario and has progressed quickly in recent years. It
provides a secure way for sharing not only a  classical information
\cite{HBB,KKI,Bandyopadhyay,Karimipour,cpyang,
longqss,circular,dengQSSentanglement,guoqss,dengsinglephoton} but
also a quantum information
\cite{HBB,KKI,cleve,Peng,dengmQSTS,dengpra}. In the latter, the
sender Alice will send an unknown quantum state to her agents, and
one of them can recover it with the help of the other
\cite{cleve,Peng,dengmQSTS,dengpra}. Now, QSS has also been studied
in experiment \cite{TZG,AMLance,QSSE1,AMLance2}. In general, QSS is
far more complex than quantum key distribution (QKD) \cite{Gisinqkd}
as the dishonest agent, a powerful eavesdropper, has a chance to
hide his eavesdropping by cheating the others.

Almost all the QSS schemes existing can be attributed to one of the
two types, the collective eavesdropping-check one and the individual
eavesdropping-check one. The feature of the QSS schemes based on a
collective eavesdropping check is that the procedure of the
eavesdropping check can be completed only when the sender Alice gets
the cooperation of the dishonest agent. That is, Alice should
require him to publish his outcomes of the measurements on the
samples chosen randomly for checking eavesdropping. Otherwise, she
and the honest agent cannot accomplish the task. The typical models
are those in the Refs.
\cite{HBB,KKI,Bandyopadhyay,Karimipour,cpyang,
longqss,dengQSSentanglement,cleve,dengpra}. In contrast, in the
individual eavesdropping-check QSS schemes, the eavesdropping check
between Alice and the honest agent does not resort to the
information published by the dishonest one. In other words, Alice
and the honest agent can analyze the error rate of their samples
independent of the dishonest one, and determine whether the quantum
channel between them is secure or not. Typical such QSS schemes are
the ones presented in Refs.
\cite{guoqss,dengsinglephoton,circular,dengmQSTS,Peng}.

There is a class of  one-way  QSS schemes whose security is based on
entanglement and a collective eavesdropping check (ECEC), called
them one-way ECEC QSS, such as those in Refs.
\cite{HBB,KKI,Bandyopadhyay,Karimipour,cpyang,longqss,TZG,QSSE1,AMLance2,AMLance,cleve,dengpra},
including the Hillery-Bu\v{z}ek-Berthiaume  (HBB) scheme \cite{HBB}
and the Karlsson-Koashi-Imoto (KKI) scheme \cite{KKI}, and the
experimental demonstrations for sharing a classical message
\cite{TZG,QSSE1,AMLance2}  or a qubit \cite{AMLance}. Although there
are differences among particular schemes, almost all of them
realized the following scenario. First, the sender Alice prepares
two photons in an entangled state $\vert \xi^\pm
\rangle_{BC}=\frac{1}{\sqrt{2}}(\vert \alpha\rangle_B\vert
\beta\rangle_C \pm \vert \bar{\alpha}\rangle_B\vert
\bar{\beta}\rangle_C)$, and sends the photon $B$ to Bob and the
photon $C$ to Charlie (here $\vert \alpha\rangle$ and $\vert
\bar{\alpha}\rangle$ are the two eigenvectors of a two-level quantum
system, so do $\vert \beta\rangle$ and $\vert \bar{\beta}\rangle$).
Secondly, Bob and Charlie choose at least two nonorthogonal
measuring bases (MBs) to measure their photons independently.
Thirdly, Alice tells Bob and Charlie which photons are chosen as the
samples for checking eavesdropping, and then the three participants
analyze the error rate of the samples collectively for determining
whether the quantum channel is secure or not. Finally, if Alice
deems that the quantum channel is secure, she and her agents distill
a private key from the outcomes for which Bob and Charlie choose two
correlated MBs. Let us emphasize here two properties of the one-way
ECEC QSS. First, the quantum systems are sent from Alice to her
agents in one way. Second, its test eavesdropping procedure depends
on the entanglement correlation of the quantum systems and the
information published by all the participants.

As the security of a QSS scheme is based on the public statistical
analysis of the error rate of the samples chosen randomly by the
three participants including the potentially dishonest agent, it
must be guaranteed against the dishonest agent with unlimited
computing power whose technology is limited only by the laws of
quantum mechanics \cite{book}. Without loss of generality, we assume
that the dishonest agent is Bob. The aim of this paper is to present
an opaque attack scheme for Bob to obtain Alice's message (a
classical one or a quantum one) without the cooperation of the other
agent Charlie, provided that quantum channel losses are high enough.
The superiority of Bob over current technology is restricted to the
possibility of near lossless photon transmission, the storage of
quantum states, and the capability of preparing and manipulating
two-photon Bell state. Same as Ref. \cite{wjk}, our scheme considers
the opportunity of eavesdropping arising due to a separation of two
procedures, namely, the eavesdropping-check procedure and the
message sharing procedure (including the one for sharing an unknown
quantum state).

Now, let us elaborate the principle of the one-way ECEC QSS with the
example of the KKI scheme \cite{KKI}. Alice encodes a randomly
binary bit string $S$, selecting for instance two sets of states
$\{\vert \psi^+\rangle, \vert \phi^-\rangle\}$ $\Leftrightarrow $ $
\{\vert 0\rangle, \vert 1\rangle \}$ and $\{ \vert \Psi^+\rangle,
\vert \Phi^-\rangle \}$ $\Leftrightarrow $ $ \{ \vert 0' \rangle,
\vert 1' \rangle \}$ which are defined as follows.
\begin{eqnarray}
\vert \psi ^{+}\rangle_{BC} &=&\frac{1}{\sqrt{2}}(\vert +z\rangle
_{B}\vert -z\rangle _{C} + \vert -z\rangle _{B}\vert +z\rangle
_{C}),\\
\vert \phi ^{-}\rangle_{BC} &=&\frac{1}{\sqrt{2}}(\vert +z\rangle
_{B}\vert +z\rangle _{C} - \vert -z\rangle _{B}\vert -z\rangle
_{C}),\\
\vert \Psi ^{+}\rangle_{BC} &=&\frac{1}{\sqrt{2}}(\vert +z\rangle
_{B}\vert +x\rangle _{C} + \vert -z\rangle _{B}\vert -x\rangle
_{C}),\\
\vert \Phi ^{-}\rangle_{BC} &=&\frac{1}{\sqrt{2}}(\vert +z\rangle
_{B}\vert -x\rangle _{C} - \vert -z\rangle _{B}\vert +x\rangle
_{C}),
\end{eqnarray}
where the subscripts $B$ and $C$ represent the two photons in an
entangled state, and $\vert \pm z\rangle$ and $\vert \pm x\rangle$
are the eigenstates of the z-spin (basis Z) and the x-spin (basis
X), respectively,
\begin{eqnarray}
\vert \pm x\rangle &=& \frac{1}{\sqrt{2}}(\vert +z\rangle \pm \vert
-z\rangle).
\end{eqnarray}
The KKI QSS scheme works with three steps below.

(1) Alice prepares two photons randomly in one of the four states
$\{ \vert \psi^+\rangle, \vert \phi^-\rangle, \vert \Psi^+\rangle,
\vert \Phi^-\rangle \}$, and sends the photon $B$ to Bob and the
photon $C$ to Charlie. The two agents Bob and Charlie locally
measure their photons  with the two bases Z and X randomly. They
repeat this procedure for obtaining an enough large set of outcomes
$S_a$ for distilling their private key $K_A=K_B \oplus K_C$. Here
$K_A$, $K_B$ and $K_C$ are the keys kept by Alice, Bob and Charlie
in secret sharing, respectively.

(2) Bob and Charlie declare a set of outcomes for a test of
eavesdropping first and then the measurement bases for all the
outcomes $S_a$. As claimed by the authors in KKI QSS scheme
\cite{KKI}, this order is very important for testing eavesdropping.
Even they invented a refined order for their QSS scheme. That is,
for the test bits, say $S_s$, the person who declared the outcome
first should be the last to declare the basis.

(3) After the information about the outcomes and the bases for the
test bits have been released, Alice tells Bob and Charlie which of
two bases the state was sent, but not which state. For completing
the error rate analysis of the test bits, Alice publishes which
states were sent for them. In this way, half of the outcomes $S_a$
are useful as Bob and Charlie choose such two correlated bases for
their measurements that they can deduce the states sent by Alice
after she announces her bases, and the other outcomes will be
discarded.

In essence, the KKI scheme \cite{KKI} is a typical model for the
one-way ECEC QSS. The HBB scheme \cite{HBB} is equivalent to the KKI
scheme if the three participants exploit the refined order to
declare the information for the test bits although the quantum
information carriers are three-particle Greenberger-Horne-Zeilinger
states.
\begin{eqnarray}
\vert G\rangle &=& \frac{1}{2}[\vert +x\rangle_A(\vert
+x\rangle_B\vert +x\rangle_C + \vert
-x\rangle_B\vert -x\rangle_C) \nonumber\\
&& + \vert -x\rangle_A (\vert +x\rangle_B\vert -x\rangle_C + \vert
-x\rangle_B\vert
+x\rangle_C) \nonumber\\
&=& \frac{1}{2}[\vert +y\rangle_A(\vert +x\rangle_B\vert -y\rangle_C
+ \vert
-x\rangle_B\vert +y\rangle_C) \nonumber\\
&& + \vert -y\rangle_A(\vert +x\rangle_B\vert +y\rangle_C + \vert
-x\rangle_B\vert -y\rangle_C),\label{GHZcorrelation}
\end{eqnarray}
where $\vert \pm y\rangle =\frac{1}{\sqrt{2}}(\vert +z\rangle \pm i
\vert -z\rangle)$. When Alice measures her photon $A$ with the basis
$X$, the two photons $B$ and $C$ are in one of the two states
$\{\vert \psi'^+\rangle, \vert \phi'^-\rangle \}$; otherwise they
are in one of the two states $\{\vert \Psi'^+\rangle, \vert
\Phi'^-\rangle \}$ if the basis Y is chosen by Alice. Here the two
bases of an entangled photon pair $BC$ are defined as
\begin{eqnarray}
\vert \psi'^+\rangle &\equiv& \frac{1}{\sqrt{2}}(\vert
+x\rangle_B\vert -x\rangle_C + \vert -x\rangle_B\vert
+x\rangle_C),\\
\vert \phi'^-\rangle &\equiv& \frac{1}{\sqrt{2}}(\vert
+x\rangle_B\vert +x\rangle_C + \vert
-x\rangle_B\vert -x\rangle_C),\\
\vert \Psi'^+\rangle &\equiv& \frac{1}{\sqrt{2}}(\vert
+x\rangle_B\vert +y\rangle_C + \vert
-x\rangle_B\vert -y\rangle_C),\\
\vert \Phi'^-\rangle &\equiv& \frac{1}{\sqrt{2}}(\vert
+x\rangle_B\vert -y\rangle_C + \vert -x\rangle_B\vert +y\rangle_C).
\end{eqnarray}
It is obvious that these two bases can be transformed into each
other by the transformations $\vert -x\rangle_C$ $\leftrightarrow$
$\vert +y\rangle_C$ and $\vert +x\rangle_C$ $\leftrightarrow$ $\vert
-y\rangle_C$, as same as the KKI scheme \cite{KKI} in which $\vert
-z\rangle_C$ $\leftrightarrow$ $\vert +x\rangle_C$ and $\vert
+z\rangle_C$ $\leftrightarrow$ $\vert -x\rangle_C$. It is not
difficult to prove that the other one-way ECEC QSS schemes
\cite{Karimipour,cpyang,longqss} are equivalent to the KKI scheme
with or without a little of modification.

%, no matter what the message sent by Alice is.

Our opaque attack scheme on the KKI scheme includes two procedures,
the one for checking eavesdropping and the one for generating a
private key (or for sharing quantum information). Let us first
describe the attack on eavesdropping check. It includes the
following two steps.

(a) The dishonest agent Bob first intercepts the photon $C$ sent
from Alice to Charlie, and replaces it with a fake photon $C'$ which
is one particle in the Bell state $\vert
\phi^+\rangle_{B'C'}=\frac{1}{\sqrt{2}}(\vert +z\rangle _{B'}\vert
+z\rangle _{C'} + \vert -z\rangle _{B'}\vert -z\rangle _{C'})$
prepared by Bob himself. Moreover, he uses a near lossless quantum
channel to receive the photons $B$ and $C$ sent from Alice, and
stores them with a quantum memory.

(b) When and only when Bob gets the information that the photon $B$
is chosen as the sample for detecting eavesdropping, he performs a
Bell-state measurement on the photons $C$ and $B'$. If he obtains
the result $\vert \phi^+ \rangle_{B'C}=\frac{1}{\sqrt{2}}(\vert
+z\rangle_{B'}\vert +z\rangle_{C} + \vert -z\rangle_{B'}\vert
-z\rangle_{C})$, Bob measures the photon $B$ with the basis $Z$ or
the basis $X$, the same as that in the original KKI scheme, and
announces the outcome of the single-photon measurement. If Bob gets
the result $\vert \psi^-\rangle_{B'C}=\frac{1}{\sqrt{2}}(\vert
+z\rangle_{B'}\vert -z\rangle_{C} - \vert -z\rangle_{B'}\vert
+z\rangle_{C})$, he first takes the operation $i\sigma_y=\vert
+z\rangle\langle -z\vert -  \vert -z\rangle\langle +z\vert$ on the
photon $B$ and then manipulates it as same as the case with the
result $\vert \phi^+ \rangle_{B'C}$. If he gets the other two Bell
states $\vert \phi^-\rangle_{B'C}=\frac{1}{\sqrt{2}}(\vert
+z\rangle_{B'}\vert +z\rangle_{C} - \vert -z\rangle_{B'}\vert
-z\rangle_{C})$ and $\vert
\psi^+\rangle_{B'C}=\frac{1}{\sqrt{2}}(\vert +z\rangle_{B'}\vert
-z\rangle_{C} + \vert -z\rangle_{B'}\vert +z\rangle_{C})$, he
declares he did not receive that particular bit. As he knows the
fact that his eavesdropping will introduce errors in the outcomes
when Bob obtains $\vert \phi^-\rangle_{B'C}$ or $\vert
\psi^+\rangle_{B'C}$, his cheating will erase the errors in the test
bits. Of course, Bob's cheating will in principle induce 50\% losses
of the outcomes in an ideal condition.

Suppose that the honest agent Charlie obtains the outcome $\vert
R_{C'}\rangle=a\vert +z\rangle + b\vert -z\rangle$ when he measures
the photon $C'$ with the basis $Z$ or $X$. The photon $B'$ will
collapse to the state $\vert R_{B'}\rangle=\vert R_{C'}\rangle$ as
well. After the Bell-state measurement was done by Bob on the
photons $B'C$  and the result $\vert \phi^+\rangle_{B'C}$ was
obtained, the photon $B$ will collapse to the the states (without
being normalized) $a\vert +z\rangle -b\vert -z\rangle$, $a\vert
-z\rangle +b\vert +z\rangle$, $(a-b)\vert +z\rangle - (a+b)\vert
+z\rangle$, and $(a+b)\vert +z\rangle + (a-b)\vert -z\rangle$ when
the entangled photon pair $BC$ is prepared by Alice in the states
$\vert \phi^-\rangle_{BC}$, $\vert \psi^+\rangle_{BC}$, $\vert
\Phi^-\rangle_{BC}$, and  $\vert \Psi^+\rangle_{BC}$, respectively.
Table I shows the states obtained by the three participants when Bob
uses our opaque attack scheme and obtains the result $\vert
\phi^+\rangle_{B'C}$. Alice's states are given in the columns and
Charlie's outcomes are given in the rows. Bob's states before he
measures the photon $B$ with a correlated basis appear in the boxes.
They are the same as those in the original KKI QSS scheme. The
difference between the results $\vert \phi^+\rangle_{B'C}$ and
$\vert \psi^-\rangle_{B'C}$ is just the unitary operation
$i\sigma_y$. That is to say, Bob's attack will introduce no error in
the test bits for detecting eavesdropping if he obtains the result
$\vert \phi^+\rangle_{B'C}$ or $\vert \psi^-\rangle_{B'C}$ no matter
what the state of the original two photons $B$ and $C$ prepared by
Alice is and the bases chosen by the two agents for their photons.
If he gets the other two Bell states, which takes place with the
probability 50\%, he will introduce about 50\% error rate in the
test bits, but he can hide these errors with cheating. That is, for
these outcomes, he tells Alice and Charlie that he gets nothing when
he measures the photon $B$ because of the quantum channel losses. So
Alice and Charlie cannot detect this eavesdropping if the losses
aroused by the quantum channel is more than 50\% no matter what the
order of the information published do the three participants choose,
not the case claimed in the original KKI QSS scheme.

\begin{table}[!h]
\begin{center}
\caption{The states obtained by the three participants Alice,
Charlie and Bob.}
\begin{tabular}{ccccccccccc}
& & & & \multicolumn{7}{c}{Alice}\\\cline{5-11}
& & & & $\vert
\phi^-\rangle $ & & $\vert \psi^+\rangle $ & & $ \vert \Phi^-\rangle
$ & & $\vert \Psi^+\rangle$ \\\cline{3-11}
         & & $\vert +z\rangle $ & \vline & $\vert +z\rangle$ & & $\vert -z\rangle$ & & $\vert -x\rangle$ & & $\vert +x\rangle$ \\
 Charlie & & $\vert -z\rangle$  & \vline & $\vert -z\rangle$ & & $\vert +z\rangle$ & & $\vert +x\rangle$ & & $\vert -x\rangle$ \\
         & & $\vert +x\rangle $ & \vline & $\vert -x\rangle$ & & $\vert +x\rangle$ & & $\vert -z\rangle$ & & $\vert +z\rangle$ \\
         & & $\vert -x\rangle $ & \vline & $\vert +x\rangle$ & & $\vert -x\rangle$ & & $\vert +z\rangle$ & & $\vert -z\rangle$%\\
\end{tabular}\label{Table2}
\end{center}
\end{table}

For the other instances used for creating a private key (not for
checking eavesdropping), Bob need only perform a Bell-state
measurement on the two photons $B$ and $C$ after Alice announces
their basis, and then he can get all the information about Alice's
key $K_A$. On the other hand, in the period of declaring bases for
the key bits, Bob can announce a fake information about the sequence
of the bases Z and X, and then measure the photon $B'$ with the same
basis as chosen by Charlie and obtain the key $K_C$ after Charlie
publishes her bases for the outcomes. In this way, the eavesdropping
done by Bob cannot be detected.

So far, we have presented our opaque attack scheme for the dishonest
agent to eavesdrop the one-way ECEC QSS fully and freely with the
typical model, the KKI QSS scheme \cite{KKI}.\emph{ This attack
scheme works for other one-way ECEC QSS schemes
\cite{HBB,Bandyopadhyay,Karimipour,cpyang,longqss,TZG,QSSE1,AMLance2,AMLance,cleve,dengpra},
including those for sharing an unknown quantum state
\cite{HBB,KKI,AMLance,cleve,dengpra}, with or without a little of
modification as Bob only measures the photons used for the test
bits, not for message before he gets all other information published
by Alice and Charlie.} Bob does not introduce errors in the test
bits but only produces losses. The losses induced by Bob can be,
however, hidden in the channel losses. Same as Ref. \cite{wjk}, let
us suppose that Alice and her agents use an original quantum channel
with a total transmission efficiency $\eta$ not exceeding 50\%.
Typical values of $\eta$ for long-distance experimental quantum
communication well fit this range \cite{wjk,nature}. For
eavesdropping, Bob replaces the original quantum channel by a better
one with the transmission efficiency $\eta'$. The probability $P_e$
that Bob can eavesdrop the one-way ECEC QSS freely with this opaque
cheating attack scheme is $P_e=\frac{2(\eta'-\eta)}{\eta'}$ when
$\eta' < 2\eta$; otherwise, $P_e=100\%$. In order to keep the
transmission efficiency $\eta$ between Alice and Charlie, Bob should
filter out $(1-\eta)\times 100\%$ of the fake photon $C'$ reaching
Charlie. In this way the information about the eavesdropping is
completely erased from the outcomes obtained by Charlie in the
detecting eavesdropping procedure and the message sharing procedure.
This means that Bob can eavesdrop fully and freely the information
in the one-way ECEC QSS if he replaces the original quantum channel
with a better quantum channel whose total transmission efficiency is
double of that of original one and controls the balance of
transmission efficiencies.

Let us now consider how to improve the one-way ECEC QSS to make it
secure. Its insecurity over a lossy quantum channel in principle
arises from two factors. One is that the dishonest agent Bob always
controls a photon which is entangled with the other one received by
the honest agent before the participants do their eavesdropping
check. The other one is that the test eavesdropping procedure is a
collective one, i.e., dependent on the information published by the
dishonest agent. The first factor gives Bob the chance that he can
obtain a correlated outcome with a certain probability in the
eavesdropping check procedure and also get correctly all the
information obtained by all the participants subsequently
(especially in the case for sharing an unknown quantum state). The
second factor provides Bob the tools to erase the errors produced by
his attack in the test eavesdropping procedure. For improving the
security of one-way ECEC QSS, the boss Alice can disentangle the two
photons sent from her to her two agents for the outcomes in the test
bits. In this way, the eavesdropping check procedures between Alice
and Bob, and Alice and Charlie are as same as the Bennett-Brassard
1984 (BB84) QKD scheme \cite{BB84} which is proven secure for
generating a private key \cite{proof}. We should confess that the
modified one-way ECEC QSS for sharing a private key, to some extent,
is not better than two BB84 QKD processes as the latter provides a
simple way for three participants to share a private key efficiently
without resorting to entanglement at the expense of exchanging a
little more classical information. As for the one-way ECEC QSS for
sharing an unknown quantum state, the two photons prepared by the
boss Alice and sent to her two agents for checking eavesdropping
should be in a product state.

In conclusion, we have presented an undetectable opaque cheat attack
scheme for the dishonest agent in one-way ECEC QSS to eavesdrop the
outcomes obtained by the other agent. This scheme works on the
realistic implementation of   one-way ECEC QSS if the quantum
channel losses cannot be ignored. The dishonest agent first
intercepts the photons sent from the sender and stores them. He
sends the other agent a fake photon in a Bell state and exploits the
quantum losses to hide his action when the fake photon is chosen for
detecting eavesdropping. As he can control the efficiency of
transmission, he can eavesdrop the QSS freely. Obviously, this
attack is efficient in the case there are more than two agents in a
QSS scheme with a little modification as well. We also suggest a
general way for improving the security of one-way ECEC QSS. It works
especially when the QSS schemes are used to share an unknown quantum
state.

This work is supported by the National Natural Science Foundation of
China under Grant Nos. 10604008 and 10435020, and Beijing Education
Committee under Grant No. XK100270454.

\textbf{Note added-} The anonymous referee thinks that there is no
security loophole in one-way ECEC  QSS schemes if a standard key
sifting procedure, in which the events where no photon was detected
are removed before the participants perform the common eavesdropping
check, is performed. This is true when the QSS schemes are used for
sharing a classical information because all the participants
measured all the particles including those used for distilling a
private key, which will give no chance for the dishonest agent to
get the outcomes obtained by the other agent perfectly. However,
this way does not work when the QSS schemes are used for sharing an
unknown quantum state as the dishonest agent need not measure the
particles for sharing the unknown quantum state (he need only
measure the particles used for checking eavesdropping), which means
that they are insecure in this time even though the participants
exploit a standard key sifting procedure to check eavesdropping.

\end{document}